# On the evaluation of spinnability of metallic materials in the powerful spinning process: a review

Un Chol Ri [1,4], Kwang Myong Kye [2], Myong Chol Pak [3]

[1] Faculty of Materials Science, **Kim Il Sung** University, Pyongyang, Democratic People's Republic of Korea

[2] Faculty of Electronics and Automation, **Kim Il Sung** University, Pyongyang, Democratic People's Republic of Korea

[3] Department of Physics, **Kim Il Sung** University, Pyongyang, Democratic People's Republic of Korea

[4] Institute of Metallic Building Materials, Pyongyang, Democratic People's Republic of Korea

**Abstract**

Currently, the rapidly developing powerful spinning processes of metals are widely used in many industrial sectors including those requiring high precision processing of metal materials, and the types and production of spun part are increasing. Evaluating the spinnability (flow formability) of material is very important for expanding the application of flow-forming process for producing a lot of products. The spinnability of metal is an important basic data that predicts defects that may appear in the processing of products in advance, makes it possible to create rational processes, and guarantees product quality. In this paper, we comprehensively analyze the data studied so far on the spinnability evaluation during powerful spinning conducted at room temperature, it was described with respect to the test methods and the theoretical methods for evaluating spinnability of metallic materials, and the effect of various factors on the spinnability.

*ruc22482@163.com

## 1 Introduction

The powerful spinning (flow forming) method, which can be divided into the shear spinning and flow spinning (tube spinning) (Fig. 1), is a plastic forming process that has a long history and is widely used to make parts with axisymmetric or hollow circular parts with a thin wall thickness [1-3].

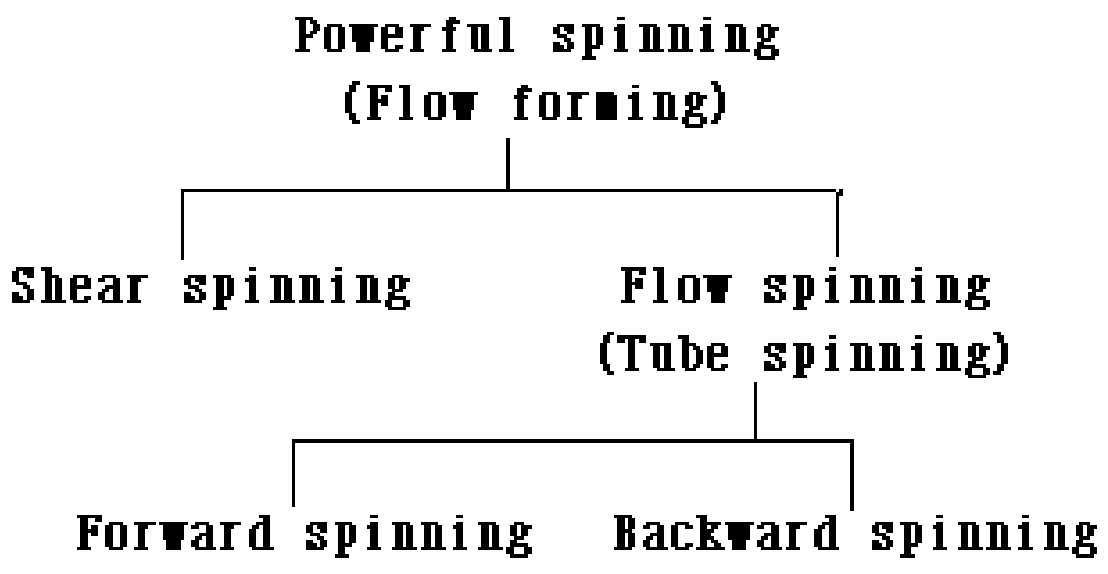

**Fig. 1** Powerful spinning classification [1-3]

As shown in Fig. 2, this method transfers one or more rollers to a metal plate material or tube material that rotates with the main axis of the machine, and gives continuous and local plastic deformation to the material to obtain the required axisymmetric and hollow products [4-7].

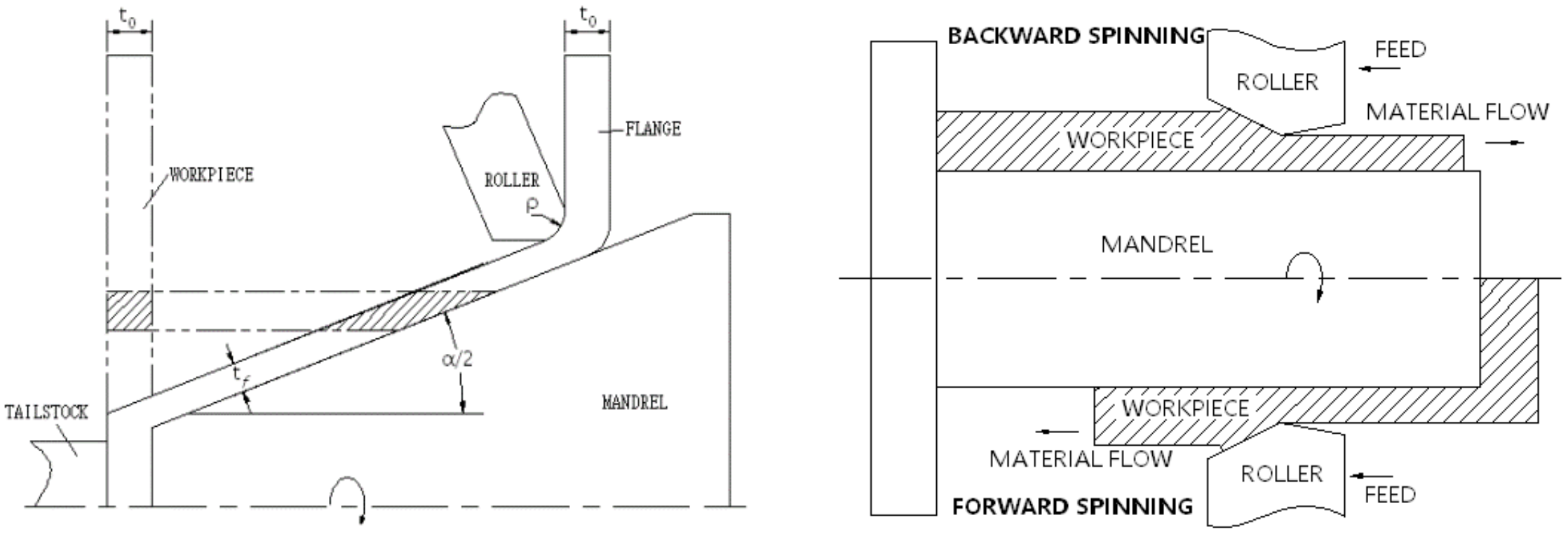

(a) (b)

**Fig. 2** Powerful spinning methods; **a** shearing spinning [4,5], **b** flow spinning [6,7]

Unlike Conventional spinning, in powerful spinning, the wall thickness decreases to obtain a product with the required forming dimensions and quality. The powerful spinning method has less force required to deform the material due to its local material deformation characteristics, and has simple tooling than other forming processes (stamping, rolling, etc.). In addition, it has the advantages of high dimensional precision, high material yield, low production cost, and improved mechanical properties. The powerful spinning method is widely used in industrial parts that require high precision processing of materials such as aerospace, aviation, automotive, medical, energy and electronics industries [8,9].

Although the powerful spinning process has been developed for a long time, there are still some problems in establishing a spinning process. One of the important problems is to predict in advance whether a material with a given dimension and properties can withstand the strain and stress applied to it without breaking. In other words, it is to predict the spinnability, which is the spinning forming ability of the material.

The forming quality of final spun is closely related to the spinnability, therefore, investigation on the spinnability is essential to improve the quality of spinning products [1,8]. In the last period, there was no test method for predicting the spinnability of metallic materials, so a lot of trial and error methods were used to establish the spinning process, and thus a lot of time and cost were consuming. From this, many studies have been conducted to evaluate the spinnability of the material.

The first approach was that Keeg developed a method of testing spinnability of various materials by conducting an elliptical mandrel experiment [10]. Then Kalpakcioglu conducted an analytical study of the problem to explain the experimental results obtained in Keeg’s spinnability tests [11]. From this analytical study, “good” and “bad” of metal spinnability were evaluated based on the true fracture deformation (0.5) during the tensile test. In addition, he proposed a test method to determine the maximum thinning rate during the forward spinning of the tube, and evaluated the spinnability of the material based on the reduction of area during tension test [3].

Meanwhile, a test method was studied to determine the maximum thinning rate in the backward spinning of the tube in which the moving direction of the roller and the workpiece were opposite, and the spinnability evaluation of the material was carried out [12]. However, this method also had the same principle as Kalpakcioglu’s spinnability test method.

In addition, a new method was proposed to more accurately evaluate the spinnability of alloys by correlating the standard tension test data of material with the flow properties of workpiece during powerful spinning, and using them together with Kalpakcioglu's spinnability test method [13]. Along with such an experimental methods, the spinnability was also studied as a theoretical methods. Hao Ma et al. compared and analyzed the tri-axial stress deviation value obtained from the tension test with the tri-axial stress deviation value obtained from the spinnability test, and presented the results of a study that the spinnability of the material can be predicted with high accuracy by the tension test data [14].

Parsa et al., which calculated the change of contact geometry by considering the plastic flow in the forward spinning process, calculated the spinnability of the tube by finite element simulation analysis [15]. Some papers proposed an improved GTN model to evaluate the spinnability of materials during multi-pass backward tube spinning, and integrated this into finite element software (ABAQUS) through a user material subroutine (VUMAT), and simulated damage evolution in several spinning passes [16,17]. The study of the finite element model to examine the spinnability took an important place in establishing the powerful spinning process of the aluminum alloy ellipsoid, and the influence of the process parameters on spinnability by analysis of variance was also considered [18].

Data on the relationship between material status before spinning and spinnability were also raised. Chang and Huang, who first defined macro and micro spinnability during tube spinning, conducted an experimental study on the spinnability for four types of aluminum alloy tubes by changing the heat treatment conditions of the material [19], and Podder et al. considered the effect of material heat treatment on the spinnability during flow spinning of AISI34 steel tube and concluded that the microscopic structure and work hardening of the material play an important role in increasing spinnability[20].

And Chun, Peng, etc. have established a spinning process using liquid forging workpiece that can reduce material consumption by accurately evaluating the spinning of aluminum alloy tubes produced by liquid forging [21], Zhang et al. obtained the result that an increase of feed rate improves the spinnability through a spinnability test of semi-continuous casting aluminum alloy tube material and introduced it into the production process [22]. And the effects of plastic deformation inhomogeneity on spinnability during the powerful spinning process were also studied experimentally in the change of several process parameters, and moderate thinning rate per pass were determined [23].

The data we have described so far are research data on cold spinnability, and the powerful spinning is not only conducted at room temperature. Alloy steels with a carbon content of 0.4% or more cannot undergo cold spinning because of poor spinnability at room temperature, therefore, these alloy steels must undergo hot spinning [24]. Some alloys with poor spinnability at room temperature, such as magnesium alloys and titanium alloys, are also undergoing hot spinning, and thus studies on hot spinnability are also being conducted [25-28].

However, this discussion of thermal spinnability is beyond the scope of this paper. Here, the spinnability was investigated for the powerful spinning that proceeds at room temperature. The studies conducted so far on the spinnability in the case of powerful spinning have become valuable basic data in establishing the spinning process of numerous metal materials. Of course, the problem of evaluating the spinnability of a material is still being studied because the stress-strain state in the deformation zone during powerful spinning is complicated. This paper synthetically analyzes the research data on the spinnability during powerful spinning, and specifically describes the spinnability and the method of spinnability test, theoretical methods for evaluating the spinnability, and the effects of various factors on the spinnability.

## 2. The spinnability of metallic material and its test methods

### 2.1 Spinnability

In general, in the metal plastic deformation process, the formability of a material is referred to as a level (amount of deformation) at which a material can be deformed before fracture occurs [29]. And the formability of the material was called in relation to the plastic working method. For example, the formability in the forging process is called forging formability, and the forging formability was evaluated with the height and width of the initial material and the deformed material [30].

Formability in powerful spinning, an advanced metal plastic deformation technology, is said to be spinnability as the ability of the workpiece to undergo spinning deformation without fracture [8,11].

Then, can metal materials that are easily deformed in plastic deformation processes such as forging, rolling, or drawing be easily deformed in the powerful spinning process? And can you see the formability in the general plastic deformation process as spinnability?

As an answer to this, the following data were investigated. Many high strength materials that would not be considered readily deformable at room temperature e.g. HS steels, surprisingly fall into the category of “good” materials, while well-known ductile materials, e.g. some aluminium alloys, appear to have intermediate properties when subjected to flow forming [13]. And castings or welding products known to have poor formability in general plastic deformation processes (rolling, drawing, stamp forging, etc.) can be plastically deformed by the powerful spinning method [19,21].

From these data, it is concluded that the spinning cannot be seen the same even though it can be viewed in relation to the formability of other plastic deformation processes. Therefore, a lot of studies related to the definition of the spinnability during c have been conducted. Kegg, who first began research on spinnability, defined spinnability as the maximum per cent reduction in thickness a material undergoes before fracture [11]. Hayama and Tago, who researched and analyzed the spinnability in the shear spinning of metal sheets, viewed the spinnability of the material as the ability to avoid wrinkles on the flange and no cracks in the blank wall thickness during the spinning process [31].

Zeng and Ma, etc. expressed the view that the spinnability is the maximum thickness reduction in percentage per pass that material could undergo before the occurrence of either buckling or failure when the deviation ratio (the presupposed wall thickness to the true wall thickness of final spun product) is zero [18].

Some researchers said that the spinnability in tube spinning process refers to the ability of a given material to undergo spinning without cracking, local instability and accumulation, etc. and given the geometric dimensions of the roller, the tube blank material and the feed rate, the maximum thinning rate is used to express the spinnability of the cylindrical part without wrinkling and cracking during each pass of spinning [9,15,32]. Therefore, spinnability is an important factor that characterizes the plastic deformation capacity of a material, indicating that the spinning characteristics are high or low, and can be expressed as the maximum thinning rate ($\psi_{\max}$) before various defects occur [22]. This Eq. (1) is cited in several literatures.

$$\psi_{\max} = \frac{t_0 - t_f}{t_0} \times 100 \text{ , \%} \tag{1}$$

where $t_0$ is the wall thickness of the workpiece, and $t_f$ is the wall thickness of the spun product. Thus, larger values of $\psi_{\max}$ mean a better material spinnability in powerful spinning [12]. If the wall thickness of the product is changed by the sine law as the shear spinning, formula(1) is expressed as follows(Eq.(2)) [11].

$$t_f = t_0 \times \sin\alpha/2 \text{ , } \psi_{\max} = (1 - \sin\alpha/2) \times 100 \text{ , \%} \tag{2}$$

However, the spinnability cannot be seen only as the maximum thinning rate per pass. Even if it is the same product and the same material in the case of powerful spinning, the maximum thinning rate at one pass is less than the total thinning rate when spinning is performed through multiple passes without intermediate heat treatment [3,33]. For example, Titanium, Inconel 825, Inconel 600, AISI304 alloys all have a maximum thinning rate of more than 70%, but this value is achieved by multi-passes (without intermediate annealing) [33]. And the maximum wall thickness reduction rate that can be allowed in one pass during powerful spinning is limited to about 80% or less, but in case of multiple passes spinning, the total thinning rate can be reduced (94~98) % without intermediate annealing [3]. The tube workpiece with thickness of 15mm can be pressed in one pass without fracture, but the thickness can be reduced to 2.9mm (thinning rate of 80.7%) by multi-passes spinning [13]. Even when material with the ultimate thinning rate of 80% is spun without intermediate annealing, this thinning rate is achieved by multiple passes rather than one pass, and the thinning rate per pass is 40% [34]. However, if the thinning rate per pass is small (usually 20%~30% or less) during the powerful spinning process of the tube, fracture being called by a variety of names such as center burst, chevron fracture etc., occurs on the inner surface of the spun product in contact with the mandrel(Fig. 3) [9,13,16,35].

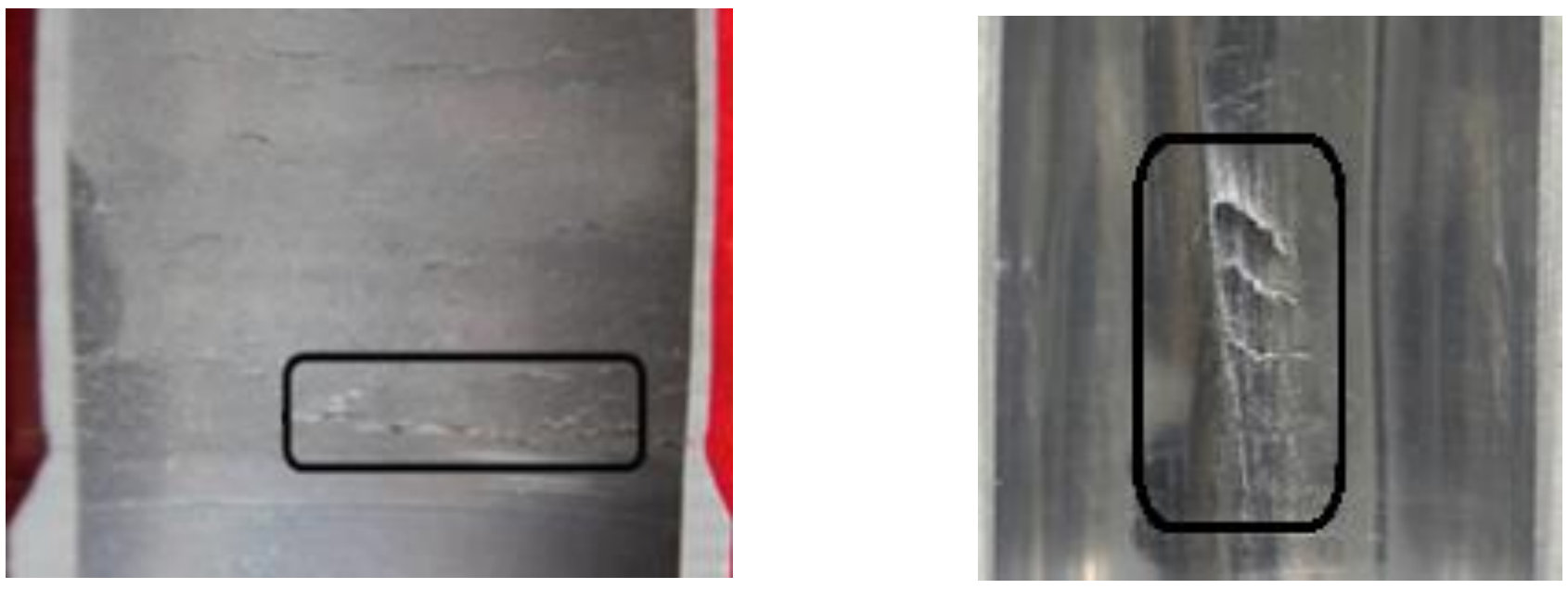

(a) (b)

**Fig. 3** Cracks occurring inside the spun product; **a** thinning rate of 18% [16], **b** total thinning rate of 24.45% (first pass of 12.35%+second pass of 13.8%) [35]

The reason is that if the thinning rate per pass is small, deformation inhomogeneity occurs between the outer layer (contact part with roller) and the inner layer (contact part with mandrel) of the spun product [23,36]. These failures is also related to the spinnability of the material, and studies have been conducted to determine the minimum thinning rate of the material per pass to prevent these defects [16,35-38]. The spinnability considered so far was evaluated based on the visually observable defects in the spinning process.

However, micro-cracks that could be observed microscopically occurred in some steels that received thinning rate more than 85%, and decreased the qualities of the product [33]. Therefore, even if no the visually observable defects at a thinning rate of 85% or more, this value cannot represent the spinnability of the material because micro-cracks decrease the strength of the spun product. From this, in the powerful spinning of 7075 and 2024 aluminum alloy, the spinnability of the material was evaluated by micro-spinnability as well as general spinnability (macro-spinnability) [19]. The micro spinnability was lower than the macro-spinnability due to the micro-cracks and micro voids present in the spun product.

Comprehensive research data show that spinnability is the ultimate ability of metallic materials to withstand spinning deformation without cracks, fractures, and other defects(Fig. 4) during the powerful spinning process [1,8-10,15,39]. Specifically, it can be said that spinnability during powerful spinning is the ultimate plastic deformation capacity of the material that can be spun (in case of multi-pass spinning, without intermediate heat treatment) without any defects (fractures, cracks, buckling, wrinkles, microscopic defects, etc.) in the workpiece under the given spinning conditions (properties of workpiece, geometry of roller, feed rate, spinning method, etc.) and is defined as the ultimate thinning rate.

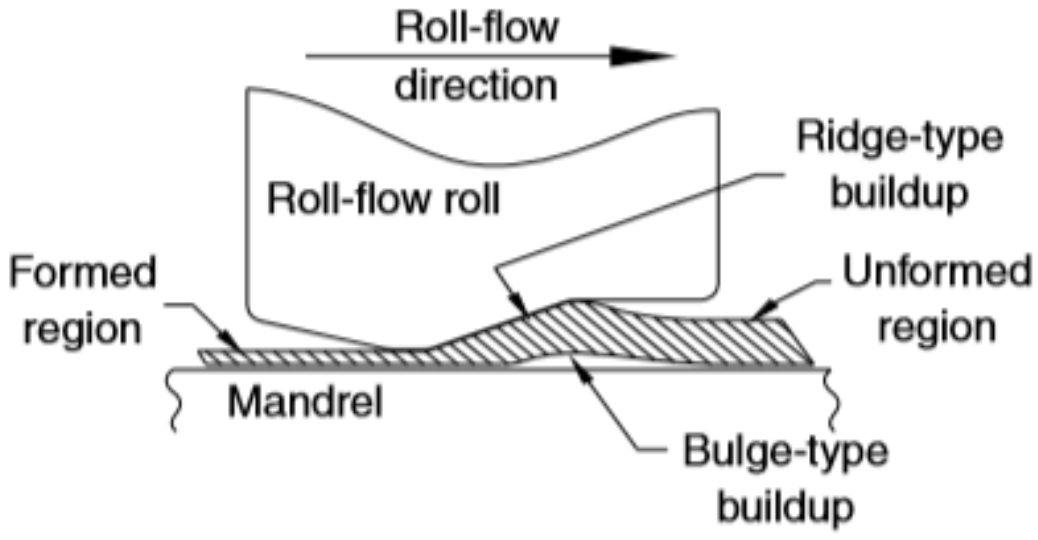

(a)

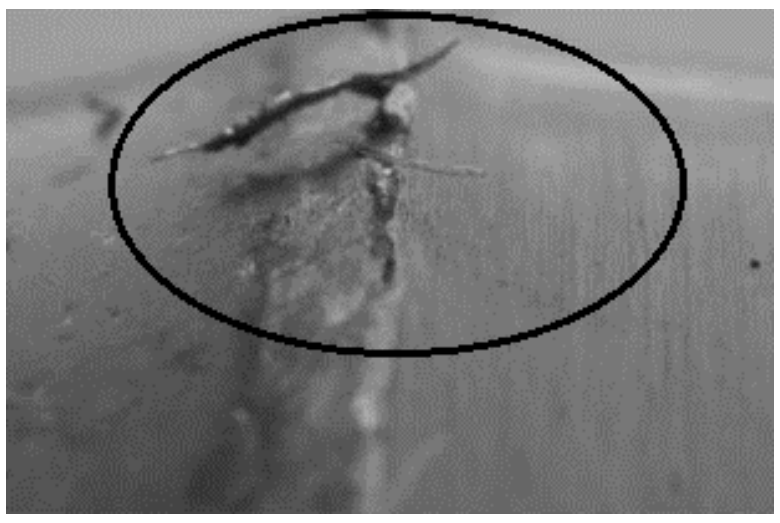

(b)

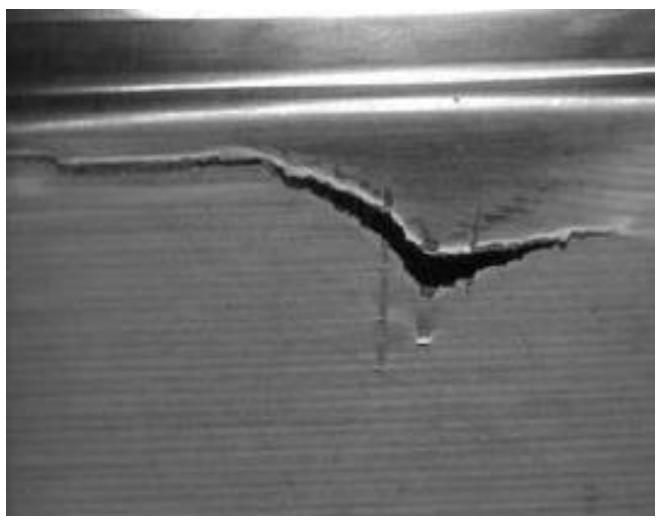

(c)

**Fig. 4 a** The type of defects occurring during powerful spinning of tube [9,39], **b** Chip formation in front of the rollers [13], **c** Penetrating fracture appearing in the spun tube [15]

### 2.2 Test methods of spinnability

In general, the purpose of testing the formability of a material in a metal plastic deformation process is to judge which material will pass which forming process through a limited number of tests [29]. For the same purpose, it can be seen that the spinnability test is also conducted in the powerful spinning. In other words, it can be seen that it tests the relationship between the properties of the material to be spun and the corresponding spinning process. The spinnability test allows for predicting in advance whether the workpiece can be spun without defects in a given process condition, and it has great significance in shortening the spinning process cycle of the spun product, increasing the productivity and lowering the production cost.

#### 2.2.1 Method for evaluating spinnability by uniaxial tensile test

Since the tube spinnability is the flow property dependent on the ductility of the material, the spinnability can be predetermined using the material index related to the ductility, such as elongation, reduction of area, and toughnes [3,10,11,13,14].

According to a study on the correlation between the shearing spinnability of metal and the tensile reduction of area, it was found that the spinnability can be predicted by knowing the tensile reduction of area of the material [11,40]. From the experimental analysis, researchers argued that the spinnability cannot be evaluated by the average elongation during the tension test and came up with an empirical formula to determine the spinnability of a material from the relationship between the spinning thinning rate and the tension reduction of area (Eq. (3)).

$$\psi_{\max} = \frac{R}{0.17 + \frac{R}{100}} \text{ , \%} \tag{3}$$

where, $R$ is tensile test reduction of area at fracture.

This formula was modified as follows for the case of the spinning of the Al alloy plate and used for the spinnability evaluation (Eq.(4)) [41].

$$\psi_{\max} = K_t \frac{R}{0.17 + \frac{R}{100}} \ , \% \qquad (4)$$

where $K_t$ is a coefficient related to the thickness of the material and has the range (0.5~1).

On the other hand, it was observed that materials with a true fracture strain of 0.5 (corresponding to 40% of reduction of area) or greater in the tension test from the spinnability test, the maximum thinning rate is not related to the ductility of the material, and spinnability of materials with a true fracture strain below 0.5 be related to the ductility of the material [11].

Similar results were also observed in the spinnability test of the tube, the spinnability (ultimate thinning rate) of materials with the tension of reduction of 45% or more was 80% or less, regardless of the ductility of the material, and the spinnability of the materials less than 45% was related to the ductility of the material [3]. Thus, the relationship between the spinnability observed in the experiment and the tension reduction of area was almost the same for the shear spinning and the tube spinning. Therefore, From this relationships one may estimate the maximum spinning reduction that during powerful process a material will withstand, knowing only the tensile property of reduction of area at fracture.

However, the stress state under the roller during the power spinning is complex, the mechanical properties obtained from simple uniaxial tensile test are insufficient to fully describe the plastic characteristics.

Also, since the relationship between the dimensions of the tension specimen and the product dimensions is not clearly defined, the obtained values cannot accurately evaluate the spinnability of the material.

Although the values obtained from the tension test are qualitative in evaluating the spinnability, they have the advantage of being able to grasp the spinnability of the material in a short time.

**2.2.2 Test method using semi-elliptical mandrel**

For the first time, the proposed test method for evaluating the spinnability of metal is a method using a semi-elliptical spinning mandrel whose semi-conical angle changes from 90° to 0° [10]. When the gap between the roller and the mandrel changes to the sine law, the wall thickness of the test workpiece gradually decreases from the initial thickness, and the half cone angle changes from a large angle to a small angle. Changes in wall thickness and semi-conical angle proceed until failure occurs in the test material (Fig.5). Spinnability is the ultimate thinning rate and is evaluated by Eq. (2). This test method, which was first studied to evaluate the spinnability of a material, has a defect in which one roller is unstable, the initial position of the roller is determined by experience, and the final thickness reduction of the test product cannot be accurately determined [18]. From this, as shown in Fig. 6 a), an improved test method for determining the spinnabillity by an elliptical mandrel was studied.

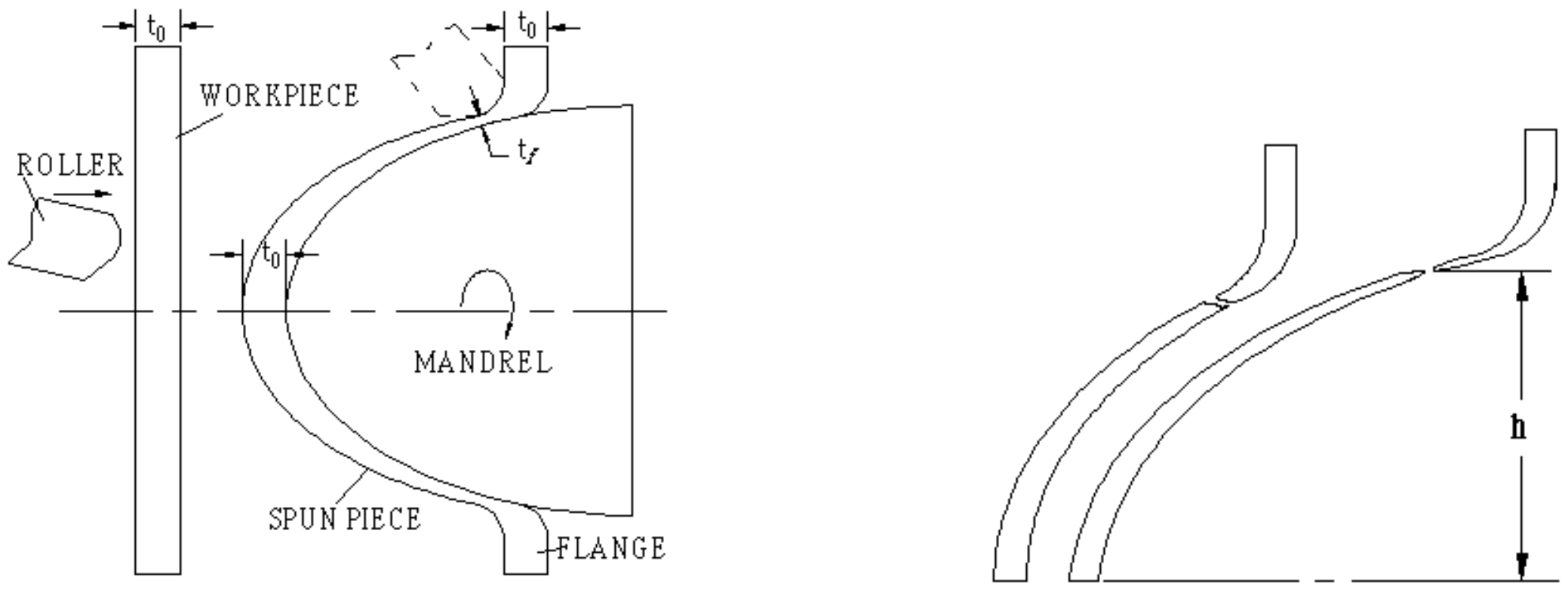

(a) (b)

**Fig. 5 a** Test method by semi-elliptical mandrel and **b** form of fractured specimen [10]

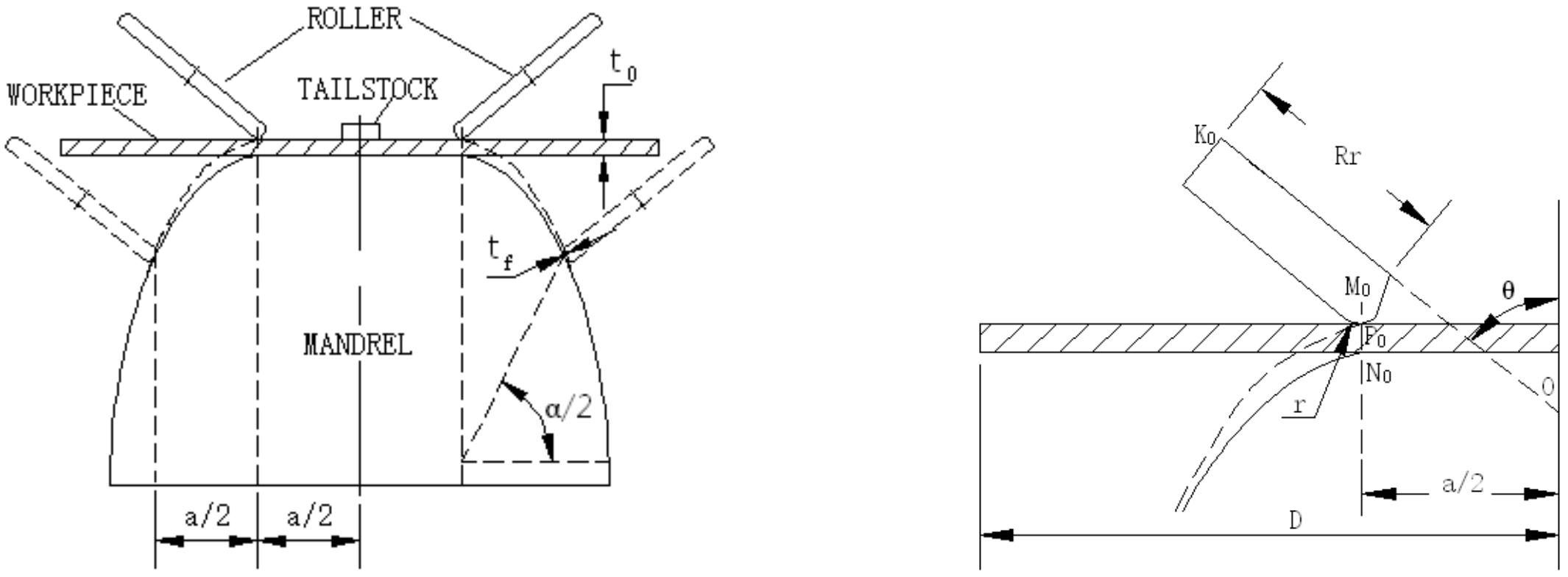

(a) (b)

**Fig. 6 a** Improved spinnability test method by elliptical mandrel and **b** Scheme for determining initial position of the roller [18]

The difference from the former is that two rollers are assembled symmetrically on both sides of the blank. In this method, the initial reference coordinate of the roller is derived from the assemble conditions (Fig. 6 b)). First, determine the initial contact point of the blank and mandrel ($N_0$), second, determine the initial fillet center point of the roller ($M_0$) and then the initial position of the roller ($K_0$) is determined by positions of these points. Two symmetric rollers are fed along the presupposed trajectory so that the ratio presupposed wall thickness to true wall thickness remain zero in theory, spinnability is evaluated by Eq. (5).

$$\psi_{\max} = \frac{t_0 - t_f}{t_0} \times 100 = (1 - \sin\frac{\alpha_{\min}}{2}) \times 100,\ \% \tag{5}$$

where $\alpha_{\min}/2$ is the minimum half ellipsoid angle.

In spinnability tests with elliptical mandrel, the wall thickness is measured at the position of the fracture point in the test product and the spinnability of the material is evaluated through the formula, and the accuracy of the wall thickness measurement value affects the test result.

Xia, etc. according to the process size parameters during spinnability test, derives formula (2) to obtain the relationship between the limit thinning rate $\psi_{\max}$ and the axial height of the spinning part h (Figure 5b)) as shown in Eq.(6) [42].

$$\psi_{\max} = \left(1 - \arctan\frac{82-h}{2\sqrt{80^2-(82-h)^2}}\right)\times 100,\ \% \tag{6}$$

By Eq. (6), the limit wall thickness can be obtained by measuring the height in the axial direction of the location where the fracture appears in the spinnability test method by elliptical mandrel. These methods are widely used in tests to evaluate the spinnability during shear spinning of conical and elliptical products.

### 2.2.3 Continuous reduction test method

The continuous reduction test method as shown in Fig. 7 is a method to test the spinnability of a material during powerful spinning of a tube [3,9,12,14,36,43].

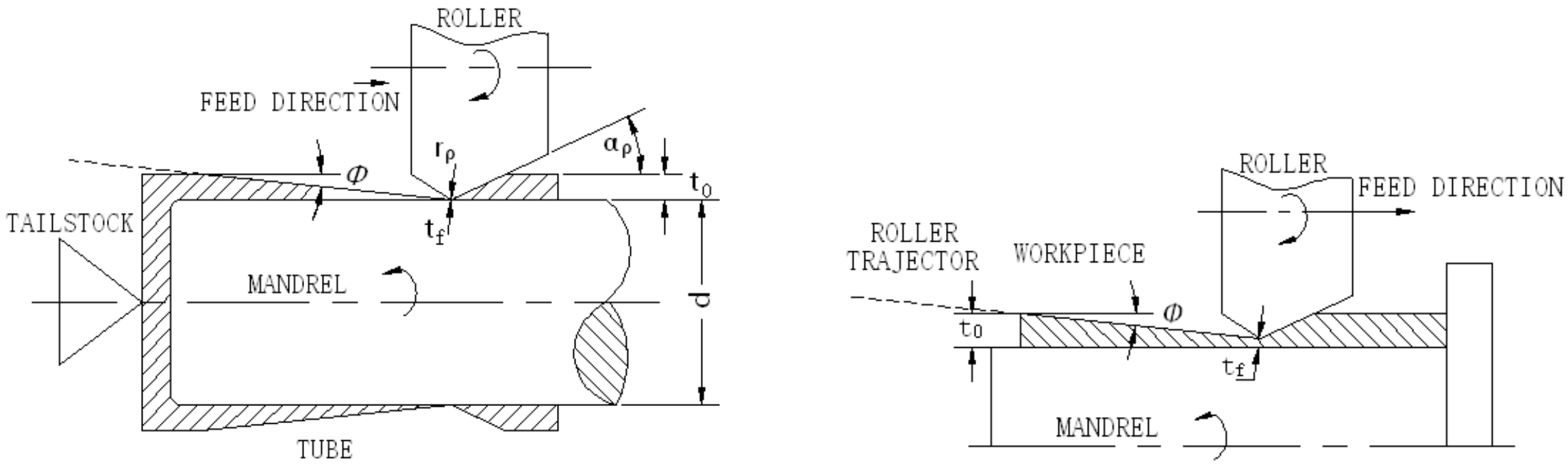

(a) (b)

**Fig. 7** The continuous reduction test method; **a** forward spinning [3], **b** backward spinning [12]

Fig. 7(a) is a test method for evaluating the spinnability during the forward spinning process, and the inner diameter (d) and thickness ($t_0$) of the used tube workpiece are 65 mm and 5 mm, respectively. The attack angle ($\alpha_\rho$) and fillet radius ($r_\rho$) of roller were 20 ° and 6mm, respectively. The mandrel rotates at 100rpm, and the rollers move along the trajectory at a small angle $\phi$ to the generatrix of the mandrel, gradually decreasing the wall thickness of the tube workpiece from the initial $t_0$ to $t_f$ (wall thickness at the location where failure occurs).

There is a limit to this one-step thinning after which the material will break in tension behind the roller. Then, after measuring the thickness at the fracture location, determine the ultimate thinning rate by Eq. (1)[3,14]. In this test, the forward spinning method in which the roller and the workpiece move in the same direction was used. Fig. 7(b) shows the spinnability test method in the backward spinning process in which the moving direction of the roller and the workpiece is opposite [12].

In the continuous reduction test method, the selection value of the wedge angle ($\phi$) has a great influence on the test result. In general, the spinnability decreases as the wedge angle increases [14]. Also, the maximum thinning rate obtained here is the limit thinning rate in one pass, so it cannot be said that the spinnability of the material is completely evaluated [9,13].

#### 2.2.4 Stepwise reduction test method

This test consists of several successive passes and in each pass, the thickness is reduced by some optimal amount (or upper) [13,19].

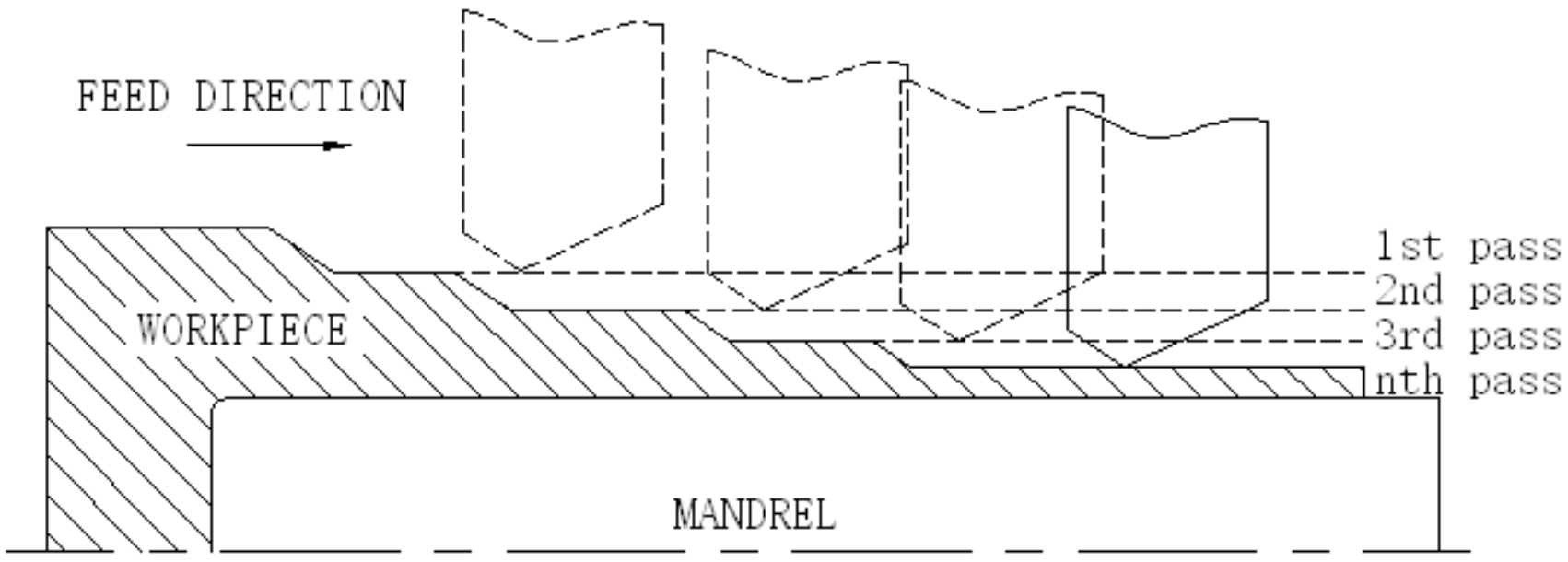

**Fig. 8** Stepwise reduction test method during forward spinning of tube [13]

Each time the rollers move along the complete length of the preform each progression being started with an axial shift forming a series of steps. Of course, in comparison with the previous tests, this test takes more passes to complete, but this started from the attempt to have a similar level of triaxiality at all stages of deformation. The results measured by the stepwise reduction test method can generally be larger than the results measured at the one-pass spinning, and there are relatively many influencing factors in the measurement process. Also, the spinnability evaluated in the stepwise reduction test is greater than spinnability obtained in the continuous reduction test. Therefore, it can be said that the lower limit of the spinnability is obtained in the continuous reduction test and the upper limit is obtained in the stepwise reduction test [35].

#### 2.2.5 Spinnability test method by microscopic observation

The test methods discussed above evaluated spinnability on the basis of visual failure or defects in the material during spinning, but this spinnability (macro spinnability) does not take into account the microscale cracks or fractures inside the spinning product. Therefore, it is evaluated to be greater than the spinnability (micro spinnability) considered with the microscale. Therefore, in evaluating the spinnability of long thin-walled spinning tube materials used in severe fatigue or corrosive environments, micro spinnability becomes more important than macro spinnability [19]. The spinnability test by observation of the microscopic structure is a method of evaluating the micro spinnability of a material by optical electron microscope (OM), scanning electron microscope (SEM), transmission electron microscope (TEM). It proceeds by finding and analyzing the microcracks and micro-voids existing in the surface layer and inside of the test workpiece. This method proceeds in combination with the stepwise feed test method [13].

Fig. 9 shows SEM and TEM photographs to evaluate the micro-spinnability of aluminum alloys. As can be seen, micro-cracks were observed in spun products that underwent the thinning rate of 70% and 50%, and spinnability of these alloys was evaluated lower than these thinning ratios. The micro-spinnability evaluated by this method will be used to establish the production process of spun products used in harsh environments and conditions such as dynamic loading and sever corrosion.

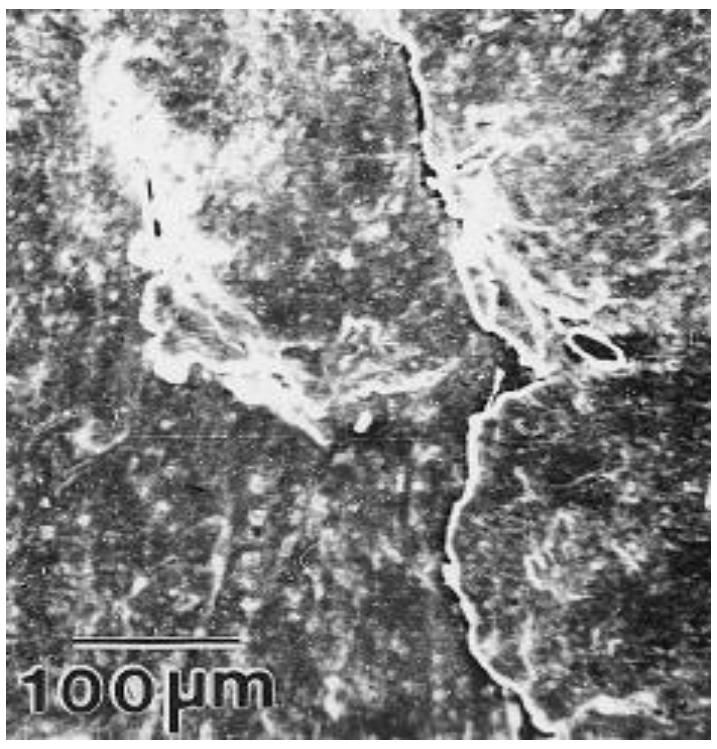

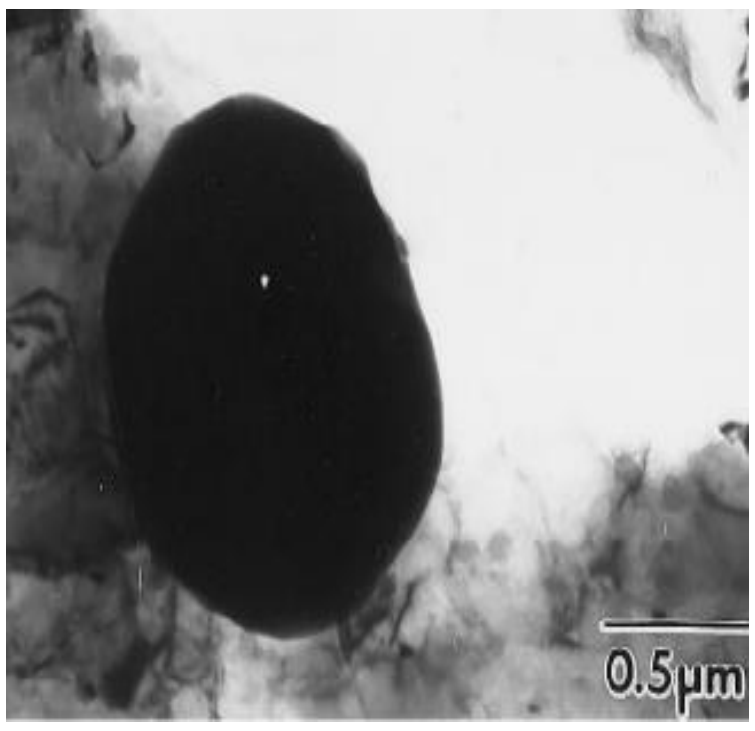

(a) (b)

**Fig. 9 a** Picture of micro-crack observed by SEM (thinning rate 70%), **b** Micrograph of particulate associated micro-crack observed by TEM (thinning rate 50%) [19]

## 3. On the theoretical methods for evaluating spinnability

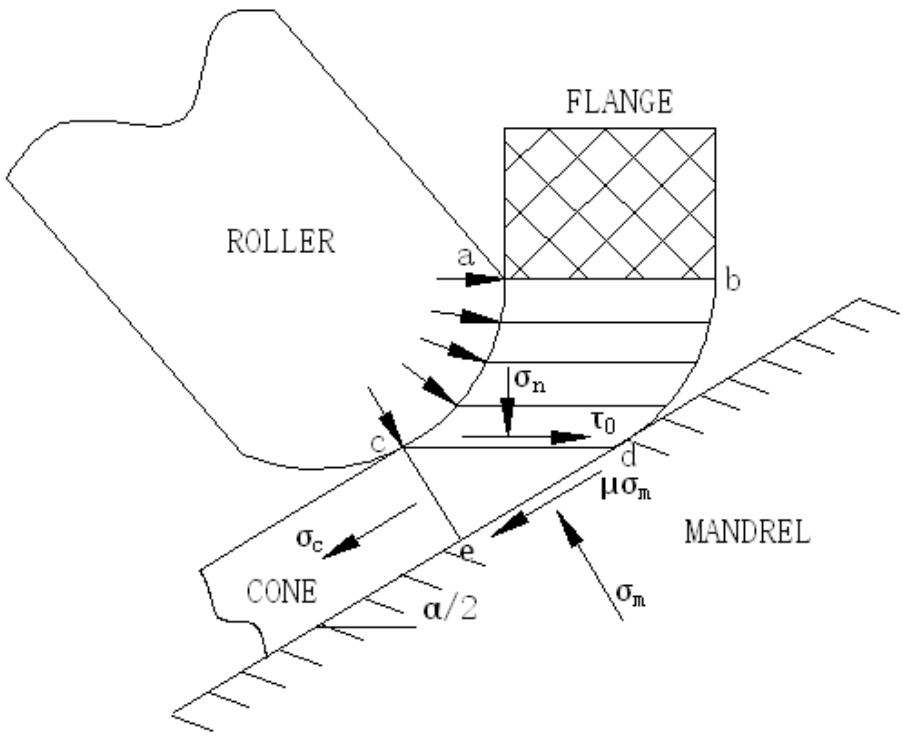

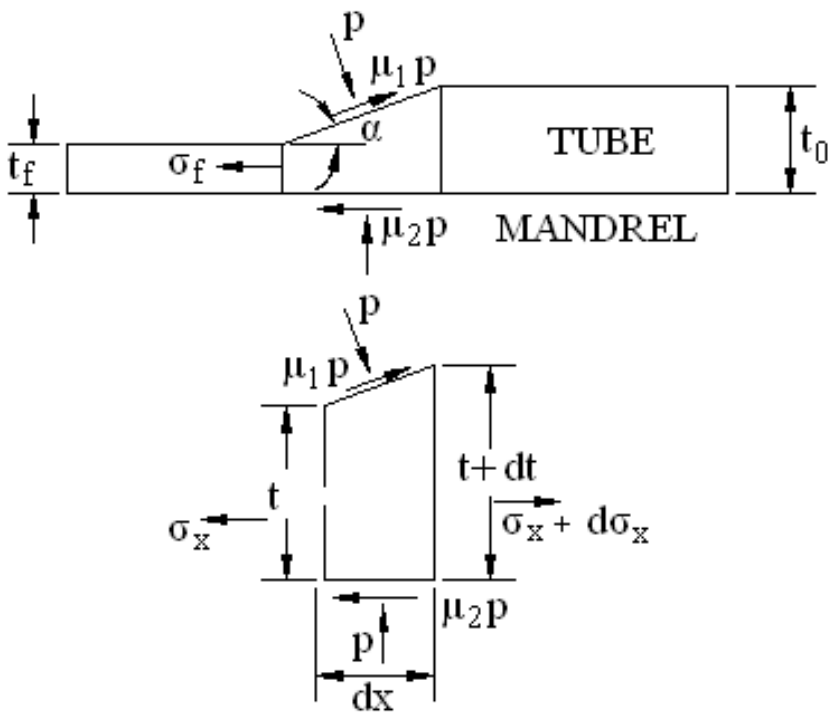

(a) (b)

**Fig. 10** Deformation zone during powerful spinning [3]; **a** cone spinning, **b** tube spinning [11]

By theoretical methods, detailed data on the stress strain in the deformation zone during powerful spinning are obtained, and by combining these data with various damage models, it is possible to predict the occurrence of defects in the spun product and to evaluate spinnability (ultimate thinning rate) under different processing conditions [1]. First, since the spinning process has complex deformation characteristics, an analytical model was prepared for an ideal material that did not consider strain hardening [11]. The stresses of the plastic zone in powerful spinning will now be analyzed on two dimensional scale as shown in Fig. 10 [3,11]. The spinnability during the shear spinning could be determined as a theoretical model (Fig. 10 a)) for analyzing the change in wall thickness and deformation characteristics of a material, it is as shown in Eq. (6) [11].

$$\alpha/2 = \cot^{-1}\left(\frac{\sigma_c}{\tau_0} + \frac{\sigma_n}{\tau_0}\right) \tag{7}$$

where, $\sigma_c$ is Normal stress acting on the plane of the deformation zone ce, $\sigma_n$ is Normal stress acting on the tangent plane cd of the deformation zone, $\tau_0$ is Shear stress in the tangential plane (determined by the base material). By substituting this $\alpha/2$ value into the Eq. (5), we can determine the value of the ultimate thinning rate. Fig. 10(b) shows the deformation zone beneath the roller assumed as a two-dimensional stress strain zone in order to determine the spinnability during the powerful forward spinning of the tube [3].

Based on this model, the force balance formula is established and the problem is simplified to derive the formula for determining the linear pressure as follows (Eq. (7)).

$$\psi_{\max} = (1 - e^{-1/k}) \times 100 \text{ , \%} \tag{7}$$

where, $k$ =1 for the ideal plastic deformable material and $k$ <1 for the strain hardening material, and is expressed as follows by the true stress-true strain relationship (Eq. (8)).

$$k = \frac{\int_0^{\varepsilon} \sigma \cdot d\varepsilon}{\sigma \cdot \varepsilon} \tag{8}$$

$k$ is often treated as a constant because the strain $\varepsilon$ is large in spinning process.

Also, since the deformation is localized within a small area and the size of the roller is large in comparison with the thickness of the tube, the contact between the roller and the tube may be considered as linear. And the simplified analysis presented herein which assumes that there is no flow in the circumferential direction and no build-up in diameter [44].

However, in the actual powerful spinning process, there is also a flow along the circumferential direction, and defects such as build-up occur. The upper bound analysis was used to theoretically analyze the defects caused by plastic flow instability (build-up, etc.) occurring at the beginning of the spinning. Here, the backward spinning process of the tube was viewed as the extrusion process and the spinning deformation process was analyzed [45]. From the comparison of the circumferential contact length (S) and the axial contact length (L) of the roller, the deformation at the powerful spinning when S>>L, it can be seen as plan extrusion deformation, and when S<<L, it can be seen as rolling deformation. Based on this material flow in the circumferential and axial directions, a mathematical model can be obtained and the occurrence of defects can be predicted by the relative S/L ratio [37,45]. The data that calculated the ratio of the circumferential contact length (S) and the axial contact length (L) of the roller and studied the linearity of the material are presented in various documents [37,45,46].

Under the assumption that the diameter of the material is constant during the shear spinning, the theoretical model for analyzing the spinnability by simplifying the complex spinning process is shown in Fig.10 [47,48]. Here, the spinning process is considered as a plane strain process, and the deformations are expressed by the diameter of the workpiece, and the spinnability is evaluated by the forming limit diagram.

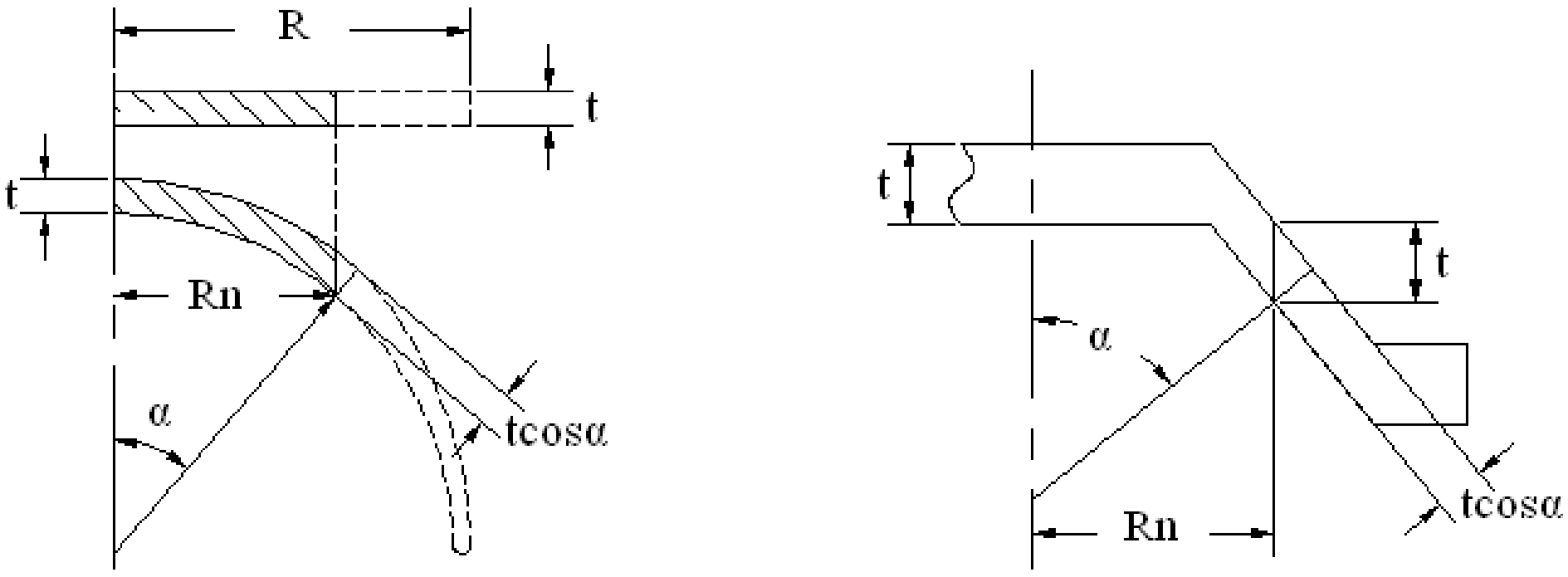

(a) (b)

**Fig. 10 a** Theoretical model of the shear spinning process [47], **b** principle diagram showing the thickness deformation [48]

However, when comparing the results obtained by this theoretical model with the experimental results, it can be seen that there are many differences [47]. As such, due to the complexity of the spinning process, the analytical studies on the spinnability were carried out by many assumptions and simplifications, so the accuracy was not high in predicting the failure in the spinning process and evaluating the spinnability.

Roy et al. have obtained the distribution of equivalent plastic deformation according to the wall thickness during powerful spinning through a micro-indentation hardness testing and analyzed the correlation with the ultimate thinning rate, and have established a method that could more accurately evaluate the spinnability and further develop finite element modeling [49].

In order to accurately evaluate the spinnability during the powerful spinning, there should be a simulation model and method that can predict the defects (failure, crack) that appear in the spinnability with high accuracy. Recently, finite element analysis methods are widely used to predict defects in the powerful spinning process, so detailed analysis data on the metal flow characteristics that cannot be obtained by analytical methods have been obtained [6,50,51].

The powerful spinning process is very difficult to model due to the following factors: repetitive contact between the roller and the workpiece, local plastic deformation, volume change due to rotation of the workpiece, and the increase in the amount of calculation and time produced by fine mesh discretisation to allow continuity of contact [52,53]. Considering these contents, many studies have been conducted to develop finite element models suitable for predicting and analyzing defects that appear during spinning deformation, and to evaluate the spinnability [54,55].

In general, static implicit FE codes are widely used in plastic forming process analysis, but in powerful spinning analysis, this is not used due to high computational cost and unguaranteed solution convergence, and dynamic explicit FE codes are widely used [15,52,56,57].

Parsae et al., which describe the advantages of the dynamic explicit FE code compared to the static implicit FE code in the spinning simulation, simulate the spinnability of the material during forward spinning of tube, and argued the validity of the simulation experiment from the comparative analysis of this simulation result and the experimental result [15]. Using the ABAQUS/Explicit module, the zhange, which simulates the powerful spinning process of the head of a large ellipsoid with varying thickness, the stress strain state of deformation area and wall thickness of the spun product were obtained in close agreement with the experimental values, and defects occurring during spinning deformation were predicted by these information [50]. Also, during forward spinning of tube the distribution of equivalent plastic deformation simulated by the dynamic explicit finite element code showed that the equivalent plastic deformation is larger in the outer layer than in the inner layer of the tube, and if this value exceeds the ultimate plastic deformation value, the outer layer occurs first [57].

Usually, cracking in the plastic deformation processes is caused by severe plastic deformation which exceeds the material forming limit. During the powerful spinning process, spinning crack could be induced by low material ductility and great thinning rates of wall thickness [58].

In some processing conditions, however, the cracks are experimentally prone to emerge at smaller thinning rates [16,23]. Therefore, in order to accurately evaluate the spinnability and establish a reasonable spinning process for each material, it is necessary to accurately predict the initiation of failure in the deformation zone. The coupling of ductile fracture criteria (DFCs) with finite element(FE) simulation have been proposed and applied to predict crack initiation, propagation, and final rupture during metal spinning forming [13,14,18,35,58-61]. The ductile fracture criteria (DFCs) can be classified into two groups: coupled DFCs that incorporates damage accumulation into the constitutive formulas and uncoupled DFCs that neglects the influence of damage on the yield surface [14]. In the split spinning considering the kinematic effects of mandrel and roller, the spinnability of aluminum alloy was studied using finite element simulation with modified Lemaitre criterion [59]. In addition, Wang et al., as a model for predicting the occurrence of defect during powerful spinning of the tube, a finite element model combined with the Gurson-Tvergaard-Needleman (GTN) damage model was made, and experimentally evaluated the spinnability [6].

Table 1 shows the finite element simulation methods combined with the ductile fracture criterion used in the spinnability evaluation.

From above data, it can be seen that the simulation results using the ductile fracture criteria (DFCs) are in a good agreement with the experimental results during the power spinning and the spinnability evaluation is also getting accurate. In addition, the results of the study show that the ductile failure criteria can be modified to suit the characteristics of the spinnability deformation, so that unpredictable initial failure phenomena, surface cracking phenomena, and damage accumulation can be obtained almost consistent with the experiment.

**Table 1** Finite element simulation methods combined with ductile fracture criteria for spinnability evaluation

| Reference (year) | Ductile Fracture Criteria (DFCs) | | | FE simulation program | Simulation results |
|---|---|---|---|---|---|
| | Group | Criteria | Formula | | |
| [61] (2009) | coupled | Lemaitre | $\Delta D = \frac{D_C}{\varepsilon_R - \varepsilon_D}\left[\frac{2}{3}(1+\nu) + 3(1-2\nu)\eta^2\right]\left[\frac{\sigma}{K}\right]^2 \Delta\varepsilon$ | ABAQUS/ EXPLICIT | Radial crack, Forming limit |
| [18] (2015) | uncoupled | Johnson-Cook | $\varepsilon_f = [D_1 + D_2 \exp(D_3\eta)][1 + D_4 \ln \dot{\varepsilon}^*][1 + D_5 T^*]$ | ABAQUS/ EXPLICIT | Fracture behaviors |
| [14] (2015) | uncoupled | Cockcroft , Latham (C-L) | $\int_0^{\bar{\varepsilon}_f^p} \sigma_1 d\bar{\varepsilon}^p = C_1$ | ABAQUS/ EXPLICIT | Forming limit, Damage evolution |
| [6] (2016) | coupled | GTN | $\phi(\sigma, \bar{\varepsilon}^p, f) = \left[\frac{\bar{\sigma}}{\sigma_0 \bar{\varepsilon}^p}\right]^2 + 2q_1 f^*(f) \cosh\left[1 - \frac{3}{2}\frac{q_2\sigma_m}{\sigma_0(\bar{\varepsilon}^p)}\right] - (1 + q_3[f^*(f)]^*)$ | ABAQUS/ EXPLICIT | Damage evolution |
| [42] (2018) | uncoupled | Modified Oyane | $\int_0^{\bar{\varepsilon}_f^p} \left[1 + A\left(1 - \frac{\sigma_m}{\sigma_1 - \sigma_3}\right)\eta\right] d\bar{\varepsilon}^p = C_2$ | ABAQUS/ EXPLICIT | Fracture prediction, Max thinning ratio |
| [58] (2018) | uncoupled | McClintock | $\int_0^{\bar{\varepsilon}_f^p} \left[\frac{\sqrt{3}}{2(1-n)} \sinh\left\{\frac{\sqrt{3}}{2(1-n)} \frac{\sigma_1 + \sigma_2}{\bar{\sigma}}\right\} + \frac{3}{4}\frac{\sigma_1 - \sigma_2}{\bar{\sigma}}\right] d\bar{\varepsilon}^p = C_3$ | ABAQUS/ EXPLICIT | Damage evolution |
| [59] (2018) | coupled | Modified Lemaitre | $\Delta D = g'(\varepsilon)\left[\frac{2}{3}(1+\nu) + 3(1-2\nu)\eta^2\right]\left[\frac{\sigma}{K}\right]^2 \Delta\varepsilon$ | ABAQUS/ EXPLICIT | Formability Limit |
| [35] (2019) | coupled | Modified GTN | $\phi(\sigma, f^*, \bar{\varepsilon}^p, D_s) = \frac{q^2}{(1-D_s^*)^2 \sigma_y^2} + 2q_1 f^* \cosh\left[-\frac{3q_2 p}{2(1-D_s^*)\sigma_y}\right] - (1 + q_1(f^*)^2)$ | ABAQUS/ EXPLICIT | Damage evolution, Cracking mechanism |

(where $D_1 \sim D_5$ are the fracture model constants, $D_C$ the critical damage value at fracture , $D_s$ shear damage , $K$ is strength coefficient, $\nu$ is Poisson ratio, $n$ is the hardening exponent, $\sigma$ is flow stress, $g'(\varepsilon)$ is a non-linearity function representing the variation slope between the damage value and the plastic strain, $\eta = \frac{\sigma_m}{\bar{\sigma}}$ is the stress triaxiality , $\bar{\sigma}$ is the Von Mises equivalent stress, $\sigma_m$ is the mean (hydrostatic) stress, $\sigma_1$ is the first principal tensile stress, $\sigma_2$ is the second principal tensile stress, $\varepsilon_D$ is the plastic strain at the onset of damage, $\varepsilon_R$ is the plastic strain at fracture, $\dot{\varepsilon}^*$ refers to the reference strain rate, $\varepsilon_f$ is the equivalent plastic fracture strain , $\bar{\varepsilon}_f^p$ is the fracture plastic strain, $\bar{\varepsilon}^p$ is the equivalent plastic strain, $\Delta\varepsilon$ is the plastic strain increment, $C_1 \sim C_3$ are the damage limits, $f^*$ is effective porosity , $f$ is void volume fraction, $q_1$, $q_2$ are adjustment parameters for the GTN model.)

## 4. Analysis of the relationship between spinnability and various factors

How spinnability is related to various process parameters, original material properties and other factors? Such understanding is essential in order to make full use of the capabilities of powerful spinning process.

### 4.1 The relationship between the shape of spun product and the spinnability

In the case of powerful spinning, the shape of the spun product affects the spinnability of metal. Under the same conditions as the product material to be spun, the spinnability differs depending on whether the shape of the spun product is a conical product, the curved generatrix product, or a tubular product [62]. The effect of the shape of the spun product on the spinnability of the metal is shown in the Table 2.

As can be seen from Table 2, the spinnability according to the product shape is similar in cylindrical and conical products, and is evaluated as low in hemispherical products for all tested metals.

### 4.2 Relationship between spinning method and spinnability

The powerful spinning method affects the spinnability of the material.

Hayama and Kudo, who studied the effects of the backward spinning and the forward spinning on the ultimate thinning rate were argued that the forward spinning has a greater reduction rate of the ultimate thinning rate than the backward spinning, and thus the deformation conditions can be given in a wider range [63]. Such data can be found in the literature that analyzed the relationship between the ultimate thinning rate and the thickness of the initial workpiece during the forward spinning and backward spinning [41].

**Table 2** Ultimate thinning rate for different materials being spun into different shaped products [62]

| Material | Grade No. | Conical Product (%) | Hemispherical Product (%) | Cylindrical Product (%) | Curved Generatrix Product (%) |
|---|---|---|---|---|---|
| Ferrous Alloys | 6434 | 70 | 50 | 75 | 50 |
| | 4130 | 75 | 50 | 75 | 55 |
| | 4340 | 70 | 50 | 75 | 50 |
| | D6AC | 70 | 50 | 75 | 50 |
| | Rene41 | 40 | 35 | 60 | 30 |
| | 321 Stainless Steel | 75 | 50 | 75 | 50 |
| | 347 Stainless Steel | 75 | 50 | 75 | 50 |
| | 17-7PH Stainless Steel | 66 | 45 | 65 | 45 |
| | 410 Stainless Steel | 60 | 50 | 65 | 50 |
| | H11 Tool Steel | 50 | 35 | 60 | 35 |
| High Temperature Material | Mo | 60 | 45 | 60 | 35 |
| Aluminum Alloys | 2014 | 50 | 40 | 70 | 50 |
| | 5086 | 65 | 50 | 60 | 50 |
| | 6061 | 75 | 50 | 75 | 50 |
| | 7075 | 65 | 50 | 75 | 50 |

The forward spinning method can increase the spinnability compared to the backward spinning method, but in some cases it is different. Backward spinning is especially suitable for the original ductility of the preform is not high enough to accommodate the tensile stress, such as the castings and welded parts [8,19]. This is because the stress strain state in the deformation zone during backward spinning is completely different from the forward spinning method. In backward spinning, the triaxial compression stress as extrusion is created in the deformation zone in front of the roller, and in the forward spinning, tensile stress is created in the area behind the roller as drawing [9,13,36]. The triaxial compression stress state is more advantageous for plastic deformation of the material than other triaxial stress states, so the backward spinning method improves the spinnability for the spinning of workpiece such as castings and welding products. When the thickness of the workpiece is thick, the multi-pass spinning method is used more than the single-pass spinning in order to increase the spinnability, the staggered spinning method (Fig. 11) in which several rollers have a certain difference in the axial and circumferential directions is also widely used [33,63,64]. Compared with the single-pass spinning, the multi-pass spinning method can increase the spinnability because it produces smaller equivalent deformation in the inner layer of the material when the wall thickness reduction rate is the same [35].

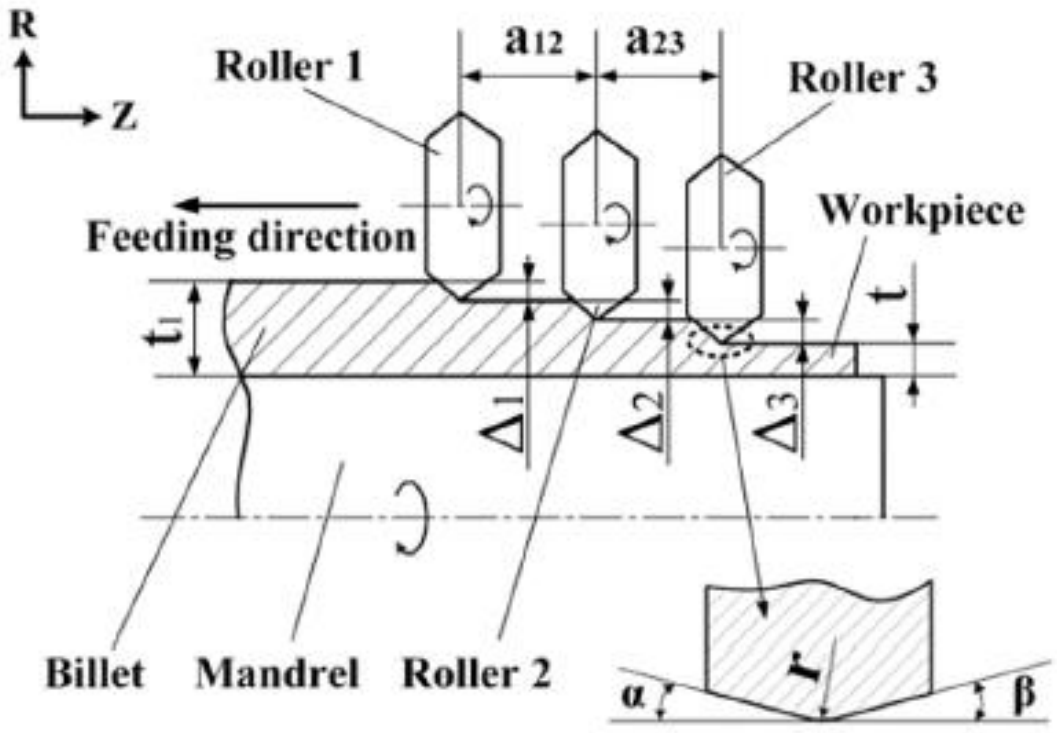

**Fig. 11** Schematic of distribution of rollers in the staggered spinning [64]

In order to further increase the spinnability in the multi-pass spinning, the wall thickness reduction rate per pass must be distributed reasonably. When the thickness of the workpiece is thick, the multi-pass spinning method is used more than the single-pass spinning in order to increase the spinnability, the staggered spinning method (Fig. 11) in which several rollers have a certain difference in the axial and circumferential directions is also widely used [33,63,64]. Compared with the single-pass spinning, the multi-pass spinning method can increase the spinnability because it produces smaller equivalent deformation in the inner layer of the material when the wall thickness reduction rate is the same [35]. In order to further increase the spinnability in the multi-pass spinning, the wall thickness reduction rate per pass must be distributed reasonably.

### 4.3 Influence of material characteristics and process parameters on spinnability

Evaluating the interactions of materials and process parameters for complex metal forming processes such as powerful spinning are very important before doing any experimental task since they define spinnability. The parameters that influence the spinnability could be divided into two major categories. One category includes the metallurgical factors such as mechanical properties of material, cleanness of alloy, and average grain size. Another category consists of mechanical factors like feed rate, roller’s fillet radius, and roller attack angle [15,19].

#### 4.3.1 Influence of material characteristics

Material characteristics are important factors that during any spinning process define the spinnability of material under given spinning condition [15]. Material characteristics that play a role are the extent and nature of impurities and inclusions, and the presence of voids prior to deformation [36]. These internal defects of the material become the source of cracking and destruction during powerful spinning, and thus lower the spinnability. Here, the mechanism of fracture appears to be based on defects generated either at inclusion-matrix interfaces or at pre-existing voids [36,46]. Although there is no visible failure by these defects during spinning, these have little effects on the tensile properties and hydraulic test which is always performed for the spun products, these defects may initiate the fatigue crack for thin-walled long tubes which undergo dynamic loading and sever corrosion environment [19]. Therefore, in order to improve the spinnability in this type of product spinning, material defects (voids, inclusions, etc.) that may cause micro-cracks or fractures must be at least removed from the material condition. To decrease effects of inclusions on the spinnability is used the refined methods such as electro-slag refined method for preparing of workpiece [46].

In addition to such materials defects, the non-uniform grain size causes the fish scaling and cracking during the powerful spinning process. The grain size of the initial material during spinning is very important in evaluating the spinnability, and the control of the grain size becomes an important problem for increasing the spinnability of the material [46,65]. The larger the grain size of the initial material, the worse the spinnability, and the finer the particle size, the better the spinnability of the material [23]. The mechanical properties of the material affect the spinnability. Spinnability of a material is associated with low flow stress, high tensile strength to yield strength ratio, high tensile elongation and reduction in area as pointed out by Kalpakjian and Rajgopa[9]. In addition, the three uniaxial elastic properties of the material obtained in the tensile test, namely elasticity, strain hardening, reduction of area are the most important in the powerful spinning process and affect the spinnability of the material [13].

However, the soft, ductile metals (e.g, aluminium alloys) have a greater tendency to form a build-up ahead of the roller, a form of defects during powerful spinning [9,46]. In the spinning process of these metals, the ultimate thinning rate must be determined so that such build-up does not occur.

The non-uniform crystal grains of the material that lowers the spinnability, and the fine grain structure and mechanical properties that increase the spinnability have much to do with the heat treatment process of the workpiece. Podder et al. studied the effect of material heat treatment on the spinnability of AISI4340 steel during powerful spinning, and pointed out that the microstructural characteristics and strain hardening properties of the workpiece obtained by heat treatment play an important role in enhancing the spinnability [20]. He proceeded with three types of heat treatment, namely, annealing, quenching + tempering, and spheroidizing for the material, and then proceeded with the spinnability evaluation test and obtained the following conclusion. That is, by the spheroidizing annealing method, a homogeneous microstructure in which secondary phases are spheroidized is obtained and the strain hardening exponent is high, so that the best spinnability of the material is obtained compared to other heat treatment methods.

In addition, through the relationship between the heat treatment and the linear pressure properties of the AISI4130 steel, it can be seen that the lower the flow stress of the initial material by heat treatment, the higher the spinnability [24]. In order to increase the spinnability of materials such as aluminum alloy or titanium alloy, which have poor formability, aging or stress relief treatment is performed [66]. The aging or stress relief treatment increases the plasticity and toughness of the material, so that materials such as aluminium alloy or titanium alloy are spun without build-up or scale. In summarizing the data, it can be seen that the more homogeneous microstructure of the material, the smaller the grain size, the smaller the amount of inclusion, the lower the flow stress, the higher the spinnability, and these properties are improved by the refining process and heat treatment process of the material. Resilience, strain hardening also affect the spinnability, and these problems must be further intensified.

### 4.3.2 Influence of spinning process parameters

spinnability is related to many process parameters during powerful spinning, and many studies have been conducted on this. The experimental analysis results for the shear spinnability showed that the process constants, such as the roller mold radius and the feed rate, did not affect the line pressure of the material, and that the minimum bending angle that could be spun from a plate to a conical product was 30 degrees [11].

On the other hand, MOHAN and MISHRA, which considered the flow characteristics of metal in the case of powerful spinning of the tube by the grid line method, pointed out that the rotational speed of the mandrel did not significantly affect the spinnability, and if the feed rate is too small or too large, surface scale defects occur due to build-up or severe strain hardening [67].

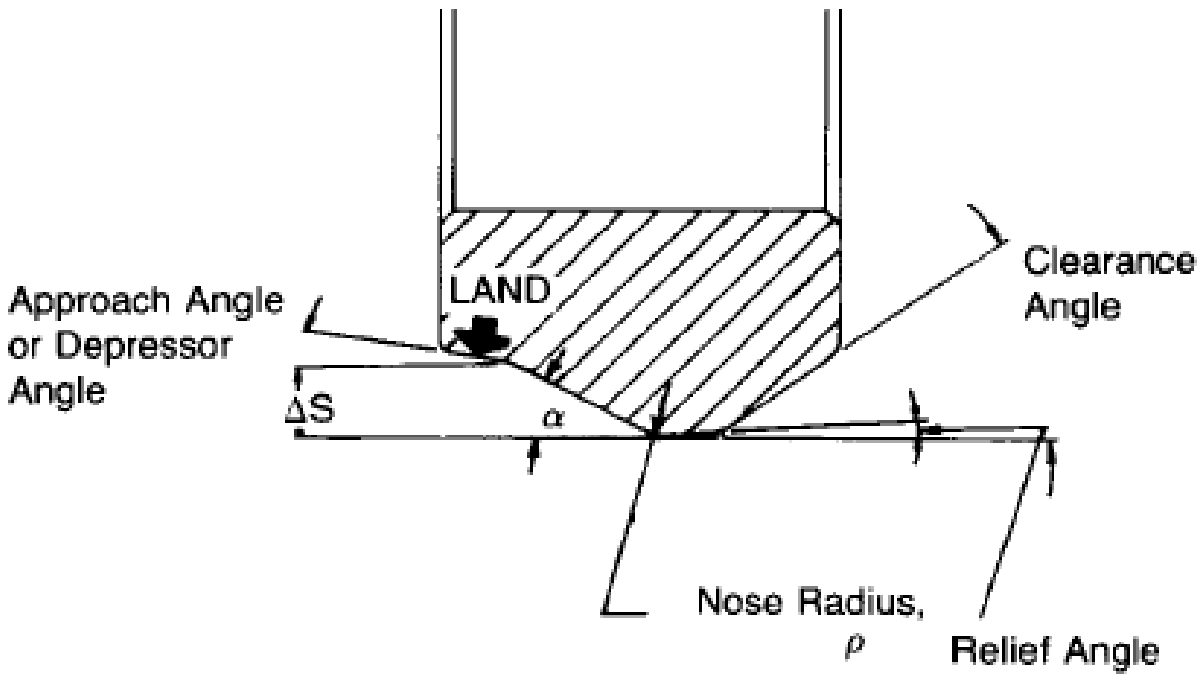

**Fig. 12** Geometry of roller to prevent build-up [9]

Defects such as scale that negatively affect the spinnability can be eliminated by changing the structure of the spinning roller as shown in the Fig. 12 [9]. In other words, by forming a depressor angle for suppressing build-up in the front of the roller, it prevents occurrence of build-up. Most of the researchers investigated the effect on the spinnability with the feed ratio defined as the ratio of the rotational speed of the mandrel and the feed speed of the roller [15,45,63,64,68].

During the forward spinning and the backward spinning of the tube, the analysis result of the relationship between the spinnability and the feed rate also showed that the smaller the feed ratio, the higher the spinnability, and the range of the roller attack angle that can increase the spinnability is (20~30) degrees [63]. In addition, according to the data that considered the plastic flow instability and process parameters causing defects in the initial stage of powerful spinning, the higher thinning rate and the smaller the feed ratio and the attack angle of the roller, the lower the probability of defects appearing [45]. This suggests that the smaller the feed ratio, the higher the ultimate thinning rate indicating spinnability.

In the paper that considered the thinning rate with the deformation characteristics of crystal grains during powerful spinning, it was pointed out that because the smaller the feed ratio, the more grains are elongated and the more broken crystal grains, the spinnability becomes higher [64]. Parsae et al., who considered the effect of the attack angle and feed speed of the roller on the spinnability, argued that by increasing feed rate, spinnability decreases and attack angle has no significant effect on spinnability from the simulation and experiment [15]. However, considering the spinning process according to the thinning rate, the roller's attack angle, and the feed ratio, it can be seen that the roller's attack angle and the ultimate thinning rate are related to each other. For example, for an attack angle of 30 degrees, the minimum reduction must be 24%, instability in plastic flow occurs and defects such as wavelike surfaces are observed below this point [37].

In the shear spinning of a conical product with inner ribs, when the feed ratio increases, the spinnability increases, and the surface quality of the spun part is poor as the feed ratio increases. Therefore, the feed ratio should be taken as large as possible under condition of obtaining acceptable surface quality in finished spun parts, thereby increasing the spinnability [68].

Under certain spinning conditions, such as low roller attack angle, low fillet of roller, and low feed rates, bulging of material ahead of the roller can also lead to scale formation and eventual cracking [2,46,63].

As discussed, process parameters have an intertwined relationship and influence the spinnability and product quality. In order to analyze the entanglement effect of process parameters on linear pressure and various objective functions, ANOVA analysis methods are widely used [18,56,69-71]. Zeng et al. evaluated the process constants that influence the spinnability during powerful spinning of the elliptical products by ANOVA analysis, and pointed out as follows. When the roller's feed speed decreases or the coefficient of friction increases, the spinnability of the elliptical product is improved, but the roller's fillet radius hardly affects the spinnability, the interaction between the roller's feed rate and the coefficient of friction has a great influence on spinnability, while the interaction of other process parameters has little effect on the spinnability [18]. According to the research data that examined the effect of process parameters on the spinnability in the single-roller backward spinning process by ANOVA analysis, the higher the roller's fillet radius and the smaller the roller's attack angle and feed ratio, the higher the spinnability [56]. On the other hand, Davidson et al. analyzed the effect of the process parameters (roller’s feed speed, mandrel rotation speed, thinning rate per pass) on the spinnability of AA6061 aluminum alloy tube by ANOVA analysis and obtained the optimal values for the maximum spinnability, and then was concluded that the thinning rate per pass had the greatest effect on the spinnability [72].

Table 3 summarizes the reviewed data on the influence of process parameters on spinnability.

**Table 3** Effect of process parameters on spinnability

| Parameters / Reference | Feed rate | Thinning rate | Geometry of roller | |
|---|---|---|---|---|
| | | | Fillet radius | Attack angle |
| [3] | Y | N | N | N |
| [11] | N | - | N | - |
| [15] | Y | - | - | N |
| [18] | Y | - | L | - |
| [35] | - | Y | - | - |
| [36] | - | Y | L | L |
| [37] | - | Y | - | Y |
| [45] | - | Y | - | Y |
| [56] | Y | - | Y | Y |
| [63] | Y | - | - | Y |
| [64] | Y | - | - | - |
| [67] | L | - | - | - |
| [68] | Y | - | - | - |
| [72] | Y | Y | - | - |

* “Y”-effect, “N”-no effect, “L”-little effect

As can be seen from the table 3, the process parameter that has the most influence on the spinnability is the feed rate (the smaller the better), and then are the thinning rate and the attack angle of the roller, and the fillet radius of the roller has a slight effect.

## 5. Conclusion

During powerful spinning, the literature related to spinnability was reviewed, and few conclusions were drawn as follows.

1) It can be said that spinnability during powerful spinning is the ultimate plastic deformation capacity of the material that can be spun (in case of multi-pass spinning, without intermediate heat treatment) without any defects (fractures, cracks, buckling, wrinkles, microscopic defects, etc.) in the workpiece under the given spinning conditions (properties of workpiece, geometry of roller, feed rate, spinning method, etc.) and is defined as the ultimate thinning rate.

2) The experimental and theoretical methods used to evaluate the spinnability have very important value in establishing the powerful spinning process of the product. In particular, the finite element simulation methods combined with the ductile fracture criterion are theoretical methods that can more quantitatively and more accurately predict all defects that appear in the powerful spinning process. In the future, an experimental method for accurately evaluating the spinnability and a method for improving theoretical models to more accurately predict defects such as fracture, damage, and cracks in the spinning process, and methods for lowering the computational cost (calculation model preparation, simulation calculation time, etc.) in the finite element simulations should be studied. In addition, methods that can reasonably combine the theoretical method and the experimental method to predict the spinnability of a material in a simple and fast time should be actively studied.

3) In order to increase the spinnability of metallic materials during powerful spinning, the following factors should be reasonably controlled. Since the spinnability of curved generatrix product is the lowest, the spinning process for such the spun product should be prepared in consideration of this relation. The melting process should be established reasonably so that metallurgical defects such as inclusions and voids do not occur in the spinning material, and the heat treatment process should be well established so that coarse grain structure or uneven grain structure does not occur. For materials(Al alloys) with good plasticity and materials (casting products and welding products, high strength alloy steels) with poor plasticity, it is necessary to select appropriate material heat treatment process, spinning method and spinning process parameters by carefully looking at the correlation between their mechanical properties (tension test data, hardening test data, compressing test data, etc.) and spinnability.

**Acknowledgments** This work was supported by the National Program on Key Science Research of Democratic People's Republic of Korea (Grant No. 20-15-32).

## References

1. Music O, Allwood JM, Kawai K (2010) A review of the mechanics of metal spinning. Journal of Materials Processing Technology 210 (1):3-23. doi:10.1016/j.jmatprotec.2009.08.021
2. Xia Q, Xiao G, Long H, Cheng X, Sheng X (2014) A review of process advancement of novel metal spinning. International Journal of Machine Tools and Manufacture 85:100-121. doi:10.1016/j.ijmachtools.2014.05.005
3. Serope K (1964) Maximum Reduction in Power Spinning of Tubes. Journal of Engineering for Industry:49-55. doi:10.1115/1.3670450
4. KALPAKCIOGLU S (1961) On the Mechanics of Shear Spinning. Journal of Engineering for Industry:125-130. doi:10.1115/1.3664441
5. Nagarajan HN, Kotrappa H, Mallanna C, Venkatesh VC (1981) Mechanics of Flow Forming. CIRP Annals 30 (1):159-162. doi:10.1016/s0007-8506(07)60915-9
6. Wang X, Zhan M, Guo J, Zhao B (2016) Evaluating the Applicability of GTN Damage Model in Forward Tube Spinning of Aluminum Alloy. Metals 6 (6). doi:10.3390/met6060136
7. Molladavoudi HR, Djavanroodi F (2010) Experimental study of thickness reduction effects on mechanical properties and spinning accuracy of aluminum 7075-O, during flow forming. The International Journal of Advanced Manufacturing Technology 52 (9-12):949-957. doi:10.1007/s00170-010-2782-4
8. Wong CC, Dean TA, Lin J (2003) A review of spinning, shear forming and flow forming processes. International Journal of Machine Tools and Manufacture 43 (14):1419-1435. doi:10.1016/s0890-6955(03)00172-x
9. KALPAKJIAN S, RAJAGOPAL S (1982) Spinning of tubes_ A review. J APPLIED METALWORKING 2:211-223. doi:10.1007/BF02834039
10. KEGG RL (1961) A New Test Method for Determination of Spinnability of Metals. Journal of Engineering for Industry:119-125. doi:10.1115/1.3664438
11. KALPAKCIOGLU S (1961) A Study of Shear-Spinnability of Metals. Transactions of the ASM E:478-484. doi:10.1115/1.3664570
12. Shi L, Xiao H, Xu DK (2017) Spinnability Investigation of High Strength Steel in Draw-spinning and Flow-spinning. Journal of Physics: Conference Series 896. doi:10.1088/1742-6596/896/1/012119
13. Bylya OI, Khismatullin T, Blackwell P, Vasin RA (2018) The effect of elasto-plastic properties of materials on their formability by flow forming. Journal of Materials Processing Technology 252:34-44. doi:10.1016/j.jmatprotec.2017.09.007

14. Ma H, Xu W, Jin BC, Shan D, Nutt SR (2015) Damage evaluation in tube spinnability test with ductile fracture criteria. International Journal of Mechanical Sciences 100:99-111. doi:10.1016/j.ijmecsci.2015.06.005
15. H. PM, A. PAM, Nili AM (2009) <15Flow-forming and flow formability simulation.pdf>. Int J Adv Manuf Technol 42:463–473. doi:10.1007/s00170-008-1624-0
16. Wu H, Xu W, Shan D, Jin BC (2019) An extended GTN model for low stress triaxiality and application in spinning forming. Journal of Materials Processing Technology 263:112-128. doi:10.1016/j.jmatprotec.2018.07.032
17. Zhan M, Gu C, Jiang Z, Hu L, Yang H (2009) Application of ductile fracture criteria in spin-forming and tube-bending processes. Computational Materials Science 47 (2):353-365. doi:10.1016/j.commatsci.2009.08.011
18. Zeng R, Ma F, Huang L, Li J (2015) Investigation on spinnability of profiled power spinning of aluminum alloy. The International Journal of Advanced Manufacturing Technology 80 (1-4):535-548. doi:10.1007/s00170-015-7025-2
19. Chang SC, Huang CA, Yu SY, Chang Y, Han WC, Shieh TS, Chung HC, Yao HT, Shyu GD, Hou HY, Wang CC, Wang WS (1998) Tube spinnability of AA 2024 and 7075 aluminum alloys. Journal of Materials Processing Technology 80-81:676-682. doi:10.1016/s0924-0136(98)00174-5
20. Podder B, Mondal C, Ramesh Kumar K, Yadav DR (2012) Effect of preform heat treatment on the flow formability and mechanical properties of AISI4340 steel. Materials & Design 37:174-181. doi:10.1016/j.matdes.2012.01.002
21. Su CW, Oon PH, Fong KS, Sina H, Wong CC (2010) Spin Formability of Al-6061 Aluminum Alloy Pre-Forms Obtained via Liquid Forging. Key Engineering Materials 447-448:422-426. doi:10.4028/www.scientific.net/KEM.447-448.422
22. Zhang RY, Yu SW, Zhang KH, Wang FC (2010) Spinnability of Semi-Continuous Casting 7A09 Aluminum Alloy. Advanced Materials Research 97-101:361-364. doi:10.4028/www.scientific.net/AMR.97-101.361
23. SHAN D-b, TONG W-z, XU Y, LV Y (2000) Effects of plastic deformation inhomogeneity on process of cold power spinning of Ti-15-3. The Chinese Journal of Nonferrous Metals 10:887-891. doi:10.19476/j.ysxb.1004.0609.2000.06.025
24. Rajana KM, Deshpandea PU, Narasimhanb K (2012) Effect of heat treatment of preform on the mechanical properties of ¯ow formed AISI 4130 Steel Tube-a theoretical and experimental assessment. Journal of Materials Processing Technology 125-126:503-511. doi:10.1016/S0924-0136(02)00305-9
25. Fata A, Tavakkoli V, Mohebbi MS (2020) Investigation of Flow-Formability of an AZ31 Magnesium Alloy. Transactions of the Indian Institute of Metals 73 (10):2601-2612. doi:10.1007/s12666-020-02047-y
26. Zhan M, Yang H, Guo J, Wang X-x (2015) Review on hot spinning for difficult-to-deform lightweight metals. Transactions of Nonferrous Metals Society of China 25 (6):1732-1743. doi:10.1016/s1003-6326(15)63778-5
27. Yoshihara S, Mac Donald B, Hasegawa T, Kawahara M, Yamamoto H (2004) Design improvement of spin forming of magnesium alloy tubes using finite element. Journal of Materials Processing Technology 153-154:816-820. doi:10.1016/j.jmatprotec.2004.04.386
28. Mohebbi MS, Rahimi pour M (2019) Effects of temperature, initial conditions, and roller path on hot spinnability of AZ31 alloy. The International Journal of Advanced Manufacturing Technology 103 (1-4):377-388. doi:10.1007/s00170-019-03528-1
29. Emmens. WC (2011) Formability_ A Review of Parameters and Processes that Control, Limit or Enhance the Formability of Sheet Metal. SpringerBriefs in Applied Sciences and Technology:1-123. doi:10.1007/978-3-642-21904-7
30. Banabic D, Bunge H-J, Pohlandt K, Tekkaya AE (2000) Formability of Metallic Materials_ Plastic Anisotropy, Formability Testing, Forming Limits. Engineering Materials:1-344. doi:10.1007/978-3-662-04013-3
31. HAYAMA M, MUROTA T, KUDO H (1966) Deformation Modes and Wrinkling Flange on Shear Spinning. Bulletin of JSME 9:423-434. doi:10.1299/jsme1958.9.423
32. ZHAO GY, ZHANG RY, GUO ZH, TU SJ, WANG EL, FENG ZR (2012) Review of Power Spinning of Thin-walled Hard-to-deform Materials Tube. Hot Working Technology 41:85-90. doi:10.14158/j.cnki.1001-3814.2012.23.043
33. SINGHAL RP, DAS SR, PRAKASH R (1987) SOME EXPERIMENTAL OBSERVATIONS IN THE SHEAR SPINNING OF LONG TUBES. Journal of Mechanical Working Technology 14:149--157. doi:10.1016/0378-3804(87)90057-X

34. Prakash R, Singhal RP (1995) Shear spinning technology for manufacture of long thin wall tubes of small bore. Journal of Materials Processing Technology doi:10.1016/0924-0136(95)01940-5
35. Wu H, Xu W, Shan D, Jin BC (2019) Mechanism of increasing spinnability by multi-pass spinning forming – Analysis of damage evolution using a modified GTN model. International Journal of Mechanical Sciences 159:1-19. doi:10.1016/j.ijmecsci.2019.05.030
36. SEROPE K (1977) Chevron Fracture in Tube Reduction by Spinning. Fracture 2:19-24. doi:10.1016/B978-0-08-022138-0.50069-1
37. Jahazi M, Ebrahimi G (2000) The influence of flow-forming parameters and microstructure on the quality of a D6ac steel. Journal of Materials Processing Technology 103:362-366. doi:10.1016/S0924-0136(00)00508-2
38. M. S, Dhami SS, Pabla BS (2012) Flow Forming Of Tubes-A Review. International Journal of Scientific & Engineering Research 3:1-11
39. Semiatin SL (2005) Metalworking _ Bulk Forming. Asm Handbook 14A:1-816
40. Hayama M (1979) Study on Spinnability of Aluminum and its Alloys. Japanese Society of Mechanical Engineers 45:1415-1425
41. SUN ZF, WANG SH (2012) Finite element numerical simulation of tube spinning. Forging & Stamping Technology 37:171-176. doi:1000-3940 (2012) 05-0171-05
42. Xia Q (2018) Ductile Fracture Criterion for Metal Shear Spinning. Journal of Mechanical Engineering 54 (1). doi:10.3901/jme.2018.14.066
43. NANJO F (2018) Spinning Workability of Al-Mg-Si Alloy. PLASTOS 1:884-885. doi:10.32277/plastos.1.12_884
44. SINGHAL RP, SAXENA PK, PRAKASH R (1990) ESTIMATION OF POWER IN THE SHEAR SPINNING OF LONG TUBES IN HARD-TO-WORK MATERIALS. Journal of Materials Processing Technology 23:29-40. doi:10.1016/0924-0136(90)90120-J
45. Gur M, Tirosh J (1982) Plastic Flow Instability Under Compressive Loading During Shear Spinning Process. Journal of Engineering for Industry 104:17-23. doi:10.1115/1.3185791
46. Beni HR, Beni YT, Biglari FR (2010) An experimental—numerical investigation of a metal spinning process. Proceedings of the Institution of Mechanical Engineers, Part C: Journal of Mechanical Engineering Science 225 (3):509-519. doi:10.1243/09544062jmes2133
47. Rajan KM, Narasimhan K (2001) An Investigation of the Development of Defects During Flow Forming of High Strength Thin Wall Steel Tubes. Journal of Failure Analysis & Prevention 1:69-76. doi:10.1007/BF02715366
48. Quigleya E, Monaghanb J (2000) Metal forming_ an analysis of spinning processes. Journal of Materials Processing Technology 103:114-119. doi:10.1016/S0924-0136(00)00394-0
49. Roy MJ, Klassen RJ, Wood JT (2009) Evolution of plastic strain during a flow forming process. Journal of Materials Processing Technology 209 (2):1018-1025. doi:10.1016/j.jmatprotec.2008.03.030
50. Zhang J, Zhan M, Yang H, Jiang Z, Han D (2012) 3D-FE modeling for power spinning of large ellipsoidal heads with variable thicknesses. Computational Materials Science 53 (1):303-313. doi:10.1016/j.commatsci.2011.08.010
51. Zhan M, Zhang T, Yang H, Li L (2016) Establishment of a thermal damage model for Ti-6Al-2Zr-1Mo-1V titanium alloy and its application in the tube rolling-spinning process. The International Journal of Advanced Manufacturing Technology 87 (5-8):1345-1357. doi:10.1007/s00170-015-8136-5
52. Wong CC, Dean TA, Lin J (2004) Incremental forming of solid cylindrical components using flow forming principles. Journal of Materials Processing Technology 153-154:60-66. doi:10.1016/j.jmatprotec.2004.04.102
53. Roy MJ, Maijer DM, Klassen RJ, Wood JT, Schost E (2010) Analytical solution of the tooling/workpiece contact interface shape during a flow forming operation. Journal of Materials Processing Technology 210 (14):1976-1985. doi:10.1016/j.jmatprotec.2010.07.011
54. Marini D, Cunningham D, Corney J (2015) A Review of Flow Forming Processes and Mechanisms. Key Engineering Materials 651-653:750-758. doi:10.4028/www.scientific.net/KEM.651-653.750
55. Podder B, Banerjee P, Ramesh Kumar K, Hui NB (2018) Flow forming of thin-walled precision shells. Sādhanā 43 (12). doi:10.1007/s12046-018-0979-7
56. Bhatt RJ, Raval HK (2018) Investigation on flow forming process using Taguchi-based grey relational analysis (GRA) through experiments and finite element analysis (FEA). Journal of the Brazilian Society of Mechanical Sciences and Engineering 40 (11). doi:10.1007/s40430-018-1456-2
57. Mohebbi MS, Akbarzadeh A (2010) Experimental study and FEM analysis of redundant strains in flow forming of tubes. Journal of Materials Processing Technology 210 (2):389-395. doi:10.1016/j.jmatprotec.2009.09.028

58. Xu W, Wu H, Ma H, Shan D (2018) Damage evolution and ductile fracture prediction during tube spinning of titanium alloy. International Journal of Mechanical Sciences 135:226-239. doi:10.1016/j.ijmecsci.2017.11.024
59. Zhan M, Guo J, Fu MW, Gao PF, Long H, Ma F (2018) Formability limits and process window based on fracture analysis of 5A02-O aluminium alloy in splitting spinning. Journal of Materials Processing Technology 257:15-32. doi:10.1016/j.jmatprotec.2018.02.021
60. TERADA K, TAKAHASHI S, TAGUCH N (2010) Mechanism and Prediction of Fracture Initiation Based onDuctile Fracture Criterion in a Spinning Draw Process. Journal of the JSTP 51:429-434. doi:10.9773/sosei.51.429
61. HU LJ, ZHAN M, HUANG L, YANG H (2009) Prediction of radial crack of the workpiece during the splitting spinning based on ductile fracture criteria. JOURNAL OF PLAST ICIT Y ENGINEERING 16:69-74. doi:10.3969/ j.issn.1007-2012.2009.03.014
62. Yang Y, Xu HJ (2010) Overview of metal spinning process. Proceedings of the 2010 IEEE International Conference on Information and Automation:2502-2508. doi:10.1109/ICINFA.2010.5512050
63. HAYAMA M, MUROTA T, KUDO H (1979) Experimental Study of Tube Spinning. Bulletin of JSME 22:769-775. doi:10.1299/kikai1938.44.3277
64. Lin YC, Qian S-S, Chen X-M, Wang J-Q, Li X-H, Yang H (2020) Influences of feed rate and wall thickness reduction on the microstructures of thin-walled Hastelloy C-276 cylindrical parts during staggered spinning. The International Journal of Advanced Manufacturing Technology 106 (9-10):3809-3821. doi:10.1007/s00170-019-04900-x
65. Shan Db, Lu Y, Li P, Xu Y (2001) Experimental study on process of cold-power spinning of Ti_15_3 alloy. Journal of Materials Processing Technology 115:380-383. doi:10.1016/S0924-0136(01)00827-5
66. PlewiŃSki A, Drenger T (2009) Spinning and flow forming hard-to-deform metal alloys. Archives of Civil and Mechanical Engineering 9 (1):101-109. doi:10.1016/s1644-9665(12)60043-0
67. Ram Mohan T, Mishra R (1972) Studies on power spinning of tubes. International Journal of Production Research 10 (4):351-364. doi:10.1080/00207547208929937
68. Ma F, Yang H, Zhan M (2010) Plastic deformation behaviors and their application in power spinning process of conical parts with transverse inner rib. Journal of Materials Processing Technology 210 (1):180-189. doi:10.1016/j.jmatprotec.2009.07.006
69. Frnčík M, Šugárová J, Šugár P, Ludrovcová B (2018) The effect of conventional metal spinning parameters on the spun-part wall thickness variation. IOP Conference Series: Materials Science and Engineering 448. doi:10.1088/1757-899x/448/1/012017
70. Nahrekhalaji ArF, Ghoreishi M, Tashnizi ES (2010) Modeling and Investigation of the Wall Thickness Changes and Process Time in Thermo-Mechanical Tube Spinning Process Using Design of Experiments. Engineering 02 (03):141-148. doi:10.4236/eng.2010.23020
71. Fazeli AR, Ghoreishi M (2010) Statistical analysis of dimensional changes in thermomechanical tube-spinning process. The International Journal of Advanced Manufacturing Technology 52 (5-8):597-607. doi:10.1007/s00170-010-2780-6
72. Davidson MJ, Balasubramanian K, Tagore GRN (2008) Experimental investigation on flow-forming of AA6061 alloy—A Taguchi approach. Journal of Materials Processing Technology 200 (1-3):283-287. doi:10.1016/j.jmatprotec.2007.09.026